\documentstyle[prl,aps,multicol,epsf]{revtex}

\input{psfig.sty}
\renewcommand{\narrowtext}{\begin{multicols}{2}
\global\columnwidth20.5pc} 
\renewcommand{\widetext}{\end{multicols}
\global\columnwidth42.5pc} \multicolsep = 8pt plus 4pt minus 3pt

\begin{document}
\draft
\title{Microscopic theory for quantum mirages in quantum corrals }
\date{\today}
\author{D. Porras$^1$, J. Fern\'{a}ndez-Rossier$^2$ and C. Tejedor$^1$}
\address{$^1$ Departamento de F\'{\i}sica Te\'{o}rica de la Materia Condensada. 
Universidad Auton\'{o}ma de Madrid. 28049 Cantoblanco, Madrid, Spain.
\\$^2$ Physics Department, University of California San Diego.
 3500 Gilman Drive, La Jolla, CA 92093. USA}
\maketitle

\begin{abstract}

Scanning tunneling microscopy permits to image the Kondo
resonance of a single  magnetic atom adsorbed on a metallic
surface. When the magnetic impurity is placed at the focus of an elliptical quantum
corral, a Kondo resonance has been recently observed both on top of the impurity
and  on top of the focus where no magnetic impurity is present.  This
projection of the Kondo resonance to a remote point on the surface
is    referred to as quantum mirage.  We present  a  quantum
mechanical  theory for  the quantum mirage inside an ideal quantum
corral and predict that the mirage  will occur in corrals with
shapes other than elliptical.

\end{abstract}


\narrowtext
\section{Introduction}

Scanning Tunneling Microscopy (STM)  allows the manipulation of
single atoms on top of a surface \cite{single} as well as the
construction of quantum structures of arbitrary shape.  Additionally, the 
differential conductance, $G(V) \equiv dI/dV$, is proportional to
the {\em local density of states} (LDOS) of the surface spot below the tip
\cite{reading}. Hence,  STM can be used to modify and to measure the
LDOS. 

A STM was used by Crommie {\em et al.} to build
a  quantum corral, {\em i.e.}, a
71 $\AA$ radius circle made with 48 atoms of iron 
 on top of a surface of copper \cite{circular}. 
  The free motion
of the electrons along the surface  changed in the presence of the 
Fe atoms so  that  quasi-bound states
appeared inside the corral.  The measured LDOS 
was quite similar to that of a gas of
noninteracting electrons inside a circular confining potential.

More recently, STM has permitted to study the problem of a
{\em single} magnetic impurity embedded in the two dimensional electron gas
formed on a metallic surface 
\cite{si1,si2}.  This is the famous Kondo problem.  Below the Kondo 
temperature, $T_K$, a many
electron singlet state forms so that the spin of the magnetic impurity
is screened by the conduction electrons.  As a consequence,
the impurity density of states develops a resonance at the Fermi energy
(the Abrikosov-Suhl resonance \cite{Hewson,Suhl}).
When the STM tip is placed on top of the magnetic impurity,
$G(V)$ displays a narrow dip around the Fermi
level \cite{si1,si2}. The  dip (instead of the resonance) is due to
 a Fano type  interference between the direct tunneling from the tip to the
surface and the additional chanels that appear due to the presence of the
impurity \cite{si1,si2,Fano,Schiller}. The
depth of the dip decreases gradually  as the lateral distance
between the tip and the impurity is increased. This
permits to image the magnetic atom. The dip vanishes  when lateral tip-magnetic impurity distance is bigger than 
$10\AA$, which is twice $k^{-1}_F$, the inverse of the Fermi vector.

This situation is dramatically changed when the magnetic impurity is
placed at the focus of an elliptical corral of size  smaller than 150
$\AA$  \cite{mirage}, built on the Cu(111) surface.
  In this configuration, the Kondo dip is
observed not only on the focus where the magnetic impurity is located
but also on the {\em empty focus}, which can be as far as 110 $\AA$ away from the
impurity.
  Remarkably, the phantom dip is not observed if either the tip or the
impurity are not at the foci. The phenomenon of the  phantom dip is
referred to as the {\em quantum mirage} \cite{mirage}.
  In this paper we provide a
quantum mechanical theory for this phenomenon. In particular  we
want to address the issue of
{\em under which conditions the quantum mirage can be
observed} and {\em whether an elliptical corral is
necessary to obtain the mirage}.  We show that the elliptical
geometry is convenient but not necessary and we show that there is no
need to invoke semiclassical arguments to explain the mirage.

The structure of this paper is the following: In section II we  
review the theoretical framework adequate to study the quantum
mirage. First, we present the Hamiltonian of a surface with both a magnetic
impurity and a quantum corral. Then we give a formal expression for
the relation between $G(V)$ and the surface LDOS.
Our original contribution starts
in section III, where we give a qualitative explanation for the
quantum mirage.  In section IV we present quantitative results for
elliptical  quantum corrals and in  section V we discuss
our results as well as the limitations of our theory.
  
  \section{Theoretical Framework}
  \subsection{The Hamiltonian of the surface}

The Hamiltonian of the surface is an extension of the well known 
Anderson model\cite{Anderson} to the case in which the electrons
feel the potential produced by the atoms creating the corral:
\begin{eqnarray}
H_{\rm surf}=\sum_{n,\sigma} \epsilon_n c_{n,\sigma}^{\dagger} c_{n,\sigma}
+ \epsilon_d \sum_{\sigma}{d_{\sigma}^{\dagger}}d_{\sigma} \nonumber \\ 
+U d^{\dagger}_{\uparrow}d_{\uparrow}d^{\dagger}_{\downarrow}d_{\downarrow} + 
V_h \sum_{n,\sigma} \psi^*_n (\vec{R}_I) 
c_{n,\sigma}^{\dagger} d_{\sigma} +h.c. . \label{Anderson}
 \end{eqnarray}
$\epsilon_n$ and $\psi_n(\vec{R})$ are the eigenvalues
and eigenfunctions of the surface corral Hamiltonian. $c_{n,\sigma}^{\dagger}$ and
$d_{\sigma}^{\dagger}$ create an electron in the state $n$ of the corral, and in the
magnetic impurity, respectively.
 In this work we only consider the
states from the metallic surface band, which seem to give the main contribution
to the LDOS measured by the STM \cite{reading,circular,mirage}. The first term in
(\ref{Anderson}) describes the Fermi sea formed by filling these states.
The second term in (\ref{Anderson}) is the impurity single particle
energy. The third term is the on-site repulsion felt
whenever two electrons are at the impurity site. The last term
describes the hopping between the surface and the impurity states.
In the Anderson model, this coupling is localized at the impurity
site, $\vec{R}_I$. From the
formal point of view, the presence of the corral is accounted for by
replacing the plane waves, which diagonalize the free surface
electron Hamiltonian, by the corral states. Throughout the paper we 
neglect the magnetic moment of the corral atoms. This is justified because
the mirage appears also when the corral atoms are non-magnetic \cite{mirage}.

 It must also be noted that
Hamiltonian (\ref{Anderson}) does not
contain any scattering from the surface states to the bulk states, a
process which could occur due to the presence of both
the impurity and the corral atoms. These physical processes should
be considered in order to have a more quantitative theory of this system, something
beyond the scope of this paper.

\subsection{$G(V,\vec{R})$ vs. LDOS}

We now review the link between  the
quantity measured in the experiments,
$G(V,\vec{R})=dI/dV(\vec{R})$,
the differential conductance measured when the tip 
is at position $\vec{R}$ on the surface, and the surface
Green's function, ${\cal G}_S(\vec{R},\epsilon^+)$. The Hamiltonian of
the whole system, tip and surface, can be written as the
sum of three terms, $H=H_{\rm tip}+H_{\rm surf}+H_{\rm tun}$. The first is
the Hamiltonian of the  tip. The second, given in equation (\ref{Anderson}),
corresponds to the Hamiltonian of the surface,
including the corral and the magnetic impurity.  The
third is the  tunneling (Bardeen) Hamiltonian, which 
describes processes in which an electron is transferred
between the tip and the surface \cite{Schiller,Chen}:
\begin{equation}
H_{\rm tun}=\sum_{\sigma} A^{\dagger}_{\sigma} 
\left(t_c \Psi_{\sigma}(\vec{R}) +
t_d d_{\sigma} \right)+h.c. ,
\end{equation}
where 
\begin{equation}
\Psi^{\dagger}_{\sigma}(\vec{R}) 
= \sum_{n} \psi^*_n(\vec{R})c^{\dagger}_{n,\sigma},
\end{equation}
creates a surface electron in the spin state 
$\sigma$ at the position $\vec{R}$ of the surface and
$A^{\dagger}_{\sigma}$ creates an electron in the tip. 
$t_c$ is the tunneling amplitude to the surface states and $t_d$ is the
amplitude for tunneling directly to the magnetic impurity. $t_d$ has to be taken
into account only when the tip is located very near the magnetic adatom
($\vec{R} \approx \vec{R}_I$). 
We assume the knowledge of the eigenstates of the tip and the
surface Hamiltonians and treat the tunneling term as a perturbation. To lowest
order in the tunneling Hamiltonian and low enough temperatures,
linear response predicts \cite{Schiller,Mahan}:
\begin{equation}
\frac{dI}{dV}(\vec{R})\equiv G(V,\vec{R})=
 \frac{4e^2}{\pi \hbar} \rho_T 
 \rho_S(\epsilon_F + eV,\vec{R}),
 \label{G}
\end{equation}
where $eV$ is the voltage drop and $\rho_T$ is the density of states of
the tip (assumed to be energy independent in the vicinity of $\epsilon_F$).
We follow the convention  that positive $eV$ means
electrons flowing towards the surface. Finally, the local density of states 
 of the surface,  $\rho_S$, is
related to the retarded surface Green's function through the relation:
\begin{equation}
 \rho_S(\vec{R},\epsilon_F + eV)=- 
 \frac{1}{\pi} Im \left[ {\cal G}_S(\vec{R},\epsilon_F + eV) \right].
\label{DOS}
\end{equation}
$\cal{G}_{S}$ is the retarded Green's function corresponding to the
operator $t_c \Psi_{\sigma}(\vec{R}) + t_d d_{\sigma} $,
and is given by \cite{Hewson,Schiller}:
\begin{eqnarray}
{\cal G}_S(\vec{R},\epsilon^{+})= 
t^2_c {\cal G}_c(\vec{R},\vec{R},\epsilon^{+}) + 
{\cal G}_d(\epsilon^{+}) \times \nonumber \\
 \left( t_d + t_c V_h {\cal G}_c(\vec{R},\vec{R}_I,\epsilon^{+}) \right)
 \left( t_d + t_c V_h {\cal G}_c(\vec{R}_I,\vec{R},\epsilon^{+}) \right),
\label{mol}
\end{eqnarray}
where $\epsilon^{+}\equiv \epsilon  + i \eta$. In the surface Green's function
(\ref{mol}), two different propagators appear. The first is the 
 impurity free ($U=0$, $V_h=0$) surface Green's function:
\begin{equation}
{\cal G}_c(\vec{R}_1,\vec{R}_2,\epsilon^{+}) =
\sum_{n}
\frac{\psi^*_n(\vec{R}_1)\psi_n(\vec{R}_2)}{\epsilon^{+}-\epsilon_n}.
\label{GC} 
\end{equation}
The second is the   Green's function at the impurity site,
 whose evaluation is the difficult part of the many body problem \cite{Hewson}.
For temperatures much lower than $T_K$, ${\cal G}_d$ can be approximated by the
Green's function of an effective resonant level with a broadening $T_K$:
\begin{equation}
{\cal G}_d(\epsilon^+)=\frac{Z_K}{\epsilon -\epsilon_F + i k_B T_K},
\label{GD}
\end{equation}
where $Z_K$ is chosen so that the impurity propagator fulfills the Friedel
sum rule \cite{Hewson}:
\begin{equation}
Z_K \approx \frac{T_K}{\pi V_h^2 \rho},
\label{Zk}
\end{equation}
where $\rho= - \frac{1}{\pi} Im \left[ {\cal{G}}_c(\vec{R}_I,\vec{R}_I,\epsilon_F)
\right]$ is the impurity free surface LDOS at the impurity site and at
$\epsilon_F$. A necessary
condition for the appearance of the Kondo resonance is that the conduction band
is formed by a quasi-continuum of states, with energy spacing $\Delta<T_K$
\cite{kondobox}. In the case of the quantum corrals
that we study below 
$\Delta>T_K$. However, the broadening, $\delta$, of these states, fulfills
$\delta>T_K$, so that the density of 
states (in the absence
of the magnetic impurity) is almost flat close to  $\epsilon_F$
and we can use equation (\ref{GD}). 
  
The surface Green's function can be expressed now as:
\begin{eqnarray}
{\cal G}_S(\vec{R},\epsilon^{+})= 
t^2_c \left( {\cal G}_c(\vec{R},\vec{R},\epsilon^{+}) + 
\frac{T_K/\pi \rho}{\epsilon-\epsilon_F +i k_B T_k} \times \right. \nonumber \\
 \left.
 \left( \frac{t_d}{t_c V_h} + 
                {\cal G}_c(\vec{R},\vec{R}_I,\epsilon^{+}) \right)
 \left( \frac{t_d}{t_c V_h} + 
                {\cal G}_c(\vec{R}_I,\vec{R},\epsilon^{+}) \right) \right).
\label{molZk}
\end{eqnarray}
For the case $t_d=0$ (tip located far from the magnetic impurity) 
we can eliminate the
parameter $V_h$ from our problem, due to the Friedel sum rule. When
the tip is placed exactly at the magnetic impurity site ($\vec{R}=\vec{R}_I$), 
$t_d$ is no longer zero and we need to estimate the parameter
$t_d/(t_c V_h)$. To do that, we proceed as follows. 
In the absence of corral atoms, we can approximate the impurity free
surface Green's function by
${\cal{G}}_c(\vec{R}_I,\vec{R}_I,\epsilon^+) \approx -i \rho_0$ and one obtains
the well known 
Fano function for the differential conductance through the tip \cite{si2,Schiller}:
\begin{equation}
G(V,\vec{R}_I) = 
\frac{4 e^2}{\pi \hbar} \rho_T \rho_0 t_c^2
\frac{(q+\epsilon')^2}{1+{\epsilon'}^2},
\end{equation}
where $\epsilon'=(eV-\epsilon_F)/T_K$, $\pi \rho_0 q =
t_d/(t_c V_h)$ and $\rho_0$ is the LDOS at the Fermi Level for
the surface states in the absence of quantum corral. $q$ is the  Fano
parameter which determines the shape of the $G(V,\vec{R}_I)$ curves. It can 
can take values between 0 (symmetric dip) and $\infty$
(Breit-Wigner). We obtain $q$, and therefore $t_d/(t_c V_h)$,
by fitting the $G(V,\vec{R}_I)$ curve to the Fano
lineshape in the case of tunneling through the magnetic adatom in the
absence of corral. 

\section{The Quantum Mirage: Qualitative Explanation}

In this section we give a qualitative explanation of the mirage, based on
the general formalism of the previous section. We need
to do several plausible hypothesis: 

We suppose that the mirage is produced by 
quasi-bound states of the corral (an assumption that is consistent
with the experiments \cite{mirage}).  
Hence, we approximate the conduction Green's function
by:
\begin{equation}
{\cal G}_c(\vec{R}_1,\vec{R}_2,\epsilon^{+}) \approx
\sum_{n}
\frac{\psi^*_n(\vec{R}_1)\psi_n(\vec{R}_2)}{\epsilon-\epsilon_n + i\delta},
\label{GC1} 
\end{equation}
where $\delta$, the broadening of the quasi-bound corral states, is roughly 20 meV
\cite{private}.

An additional approximation 
can be done if any of the two following statements holds:
\begin{itemize}
\item[(1)] The level spacing between the energies of the quasi-bound states is
much bigger than $\delta$.
\item[(2)] The level spacing is lower than $\delta$ but, due to the geometry of
the quantum corral, only a few of the bound wavefunctions take a
non-negligible value at the magnetic impurity site, $\vec{R}_I$.
If the energy separation of
these states is bigger than $\delta$, then only one of these states will
transmit the quantum mirage, as it is evident from equation (\ref{GC1}). This
condition is fulfilled in the case of the elliptic corral, as we will show below. 
\end{itemize}
In any of these two situations, whenever there is a quasi-bound state which
simultaneously has an 
energy near $\epsilon_F$ and a non-negligible density at $\vec{R}_I$, we can
replace equation (\ref{GC1}) in (\ref{molZk}) by:
\begin{equation}
{\cal G}_c(\vec{R},\vec{R}_I,\epsilon^{+}) \approx
\frac{\psi^*_{\epsilon_F}(\vec{R})\psi_{\epsilon_F}(\vec{R}_I)}
{\epsilon-\epsilon_F + i\delta}. \label{GC2}
 \end{equation}
In next section we shall use the complete expression (\ref{GC1}) for our
calculations.
 
Our last approximation is to assume $t_d << t_c$.
A finite $t_d$ is considered in next section.

When we put together all these approximations, the change in
$G(V,\vec{R})$ due to the presence of the impurity in $\vec{R}_I$ reads:
 \begin{eqnarray}
 \delta G(V,\vec{R}) \approx -\frac{4e^2V^2_ht^2_c}{\pi^2 \hbar}
  \rho_T|\psi_{\epsilon_F}(\vec{R})|^2
 |\psi_{\epsilon_F}(\vec{R}_I)|^2 \times \nonumber \\
 Im\left( \frac{1}{(eV+i \delta)^2 }\frac{1}{eV+ i k_B T_K}\right).
 \end{eqnarray}
For $eV <<\delta$ we can write:
 \begin{eqnarray}
 \delta G(V,\vec{R}) \propto  
  - |\psi_{\epsilon_F}(\vec{R})|^2
 |\psi_{\epsilon_F}(\vec{R}_I)|^2 \frac{k_BT_K}{(eV)^2+ (k_B T_K)^2}.
 \label{mirage}
 \end{eqnarray} 
 Equation (\ref{mirage}) is the most important  result of this section. We
 want to highlight several points:

{\em i)} The spectral change in $G(V)$ is a {\em dip} of width $k_BT_K$
 centered around $eV=0$, as observed in the experiments \cite{si1,si2,mirage}.

{\em ii)} According to equation (\ref{mirage}), the  dip is projected
to any point $\vec{R}$ of the corral with an strength given  by
$|\psi_{\epsilon_F}(\vec{R})|^2 |\psi_{\epsilon_F}(\vec{R}_I)|^2$.
Therefore, the projection is magnified when both the impurity and the
tip are at points where the  Fermi level corral wavefunction peaks.
The projection disappears when either the impurity or the observation
point, are located in a minimum. As we show in the next section, 
for the eccentricity of the experiment \cite{mirage} the
wavefunction of the elliptical corral at the Fermi level has its
maxima close, but not {\em at} the foci. This result is in agreement 
with the experimental observation, but reduces the importance of the role played by
  the foci.

{\em iii}) The wavefunction at the Fermi level of any quantum corral
has several maxima so that we predict that the mirage can be
observed in other geometries. A possible candidate is the stadium corral
shown in figure 3 of reference \cite{Heller}. Therefore,
{\em an elliptical geometry is not needed to observe the mirage}.

To conclude this section, we compare equation (\ref{mirage}), valid
for a confined geometry, with the case of an impurity in a
translationally invariant surface. In both cases the surface Green's
function, ${\cal G}_s$, is the sum of two contributions, the impurity
free contribution, ${\cal G}_c$, and the scattering contribution (see equation
(\ref{mol})).  The first accounts for the paths in which the electron
does not interact with the impurity and the second accounts for the
paths in which the electron does indeed interact with the impurity.
Hence, {\em the local density of states in any point of the surface
contains information about the impurity}.

 In the case of the free surface (without corral), a continuum of
quantum states with different $\vec{k}$ carries that information so
that  destructive interference takes  place at distances of the order
of $2k_F^{-1}$, the inverse of the Fermi vector \cite{Hewson}.
 In contrast, when the
electrons interact with the corral atoms, the information is carried,
essentially, by a few  quantum states, so that the destructive interference is
less efficient.  Equation (\ref{mirage})  is derived assuming that a
single quantum state is carrying the information so that there is no
interference at all.

\section{The mirage in the ellipse}

In this section we study the mirage in an elliptical corral.
Following the ideas of the previous section, we model the Green's
function of the surface states by that of the electrons confined in
an hard wall elliptical corral.
In order to compare with experiment \cite{mirage}, we consider the case in which the
corral is built on a Cu(111) surface. 
We replace the real eigenvalues of
the corral, $\epsilon_n$, by $\epsilon_n - i \delta$, in order to model
the inelastic processes, such as scattering to the bulk states.
It turns out that the problem of a quantum particle confined 
in an ellipse can be solved {\em analytically }. To do that, we write the
Schr\"odinger equation in elliptical coordinates:
\begin{eqnarray}
x&=&ae Cos[\theta] Cosh[\eta] \\
y&=&ae Sin[\theta] Sinh[\eta],
\end{eqnarray}
where $a$ and $e$ are the semimajor axis and eccentricity, respectively.
The Helmholtz equation in this coordinate system is separable, so that the
eigenstates of the problem can be written as:
\begin{equation}
\psi(\theta,\eta) = \Theta(\theta) \Lambda(\eta).
\end{equation}
The Schr\"odinger equation is written as
\begin{eqnarray}
\frac{d^2 \Lambda(\eta)}{d \eta^2} &-&(\alpha - 2 k Cosh[ 2 \eta])
\Lambda(\eta) =0 
\nonumber \\ 
\frac{d^2 \Theta(\theta)}{d \theta^2} &+&(\alpha - 2 k Cos[ 2 \theta])
\Theta(\theta) =0 \nonumber
\\
k&=&\frac{ m^* (e a)^2 \epsilon}{2\hbar^2} , 
\end{eqnarray}
where $\alpha$ is the separation constant, $\epsilon$ is the particle energy and 
$m^{*}$ is the electron effective mass which, in the Cu(111) surface band is $0.38$
$m_e$
\cite{reading,circular}.
For a given $k$, only a discrete set of $\alpha_r(k)$ meet the
requirement $\Theta(\theta)=\Theta(\theta+ 2 pi)$.  The elliptical
hard wall condition reads $\Lambda(\eta_0)=0$. It is clear that
$\eta=\eta_0$ defines an ellipse of eccentricity
$e=(Cosh[\eta_0])^{(-1)}$. For each $\alpha_r(k)$ there is a discrete
number of $k_n$ compatible with the hard wall boundary condition.
With all this in mind, we find  2 types of physically possible
solutions for the particle  inside the hard wall ellipse:
\begin{eqnarray}
\psi_{n,c}(\theta,\eta) &=&ce_r(k^c_n,\theta) Ce_r(k^c_n,\eta) \nonumber \\
\psi_{n,s}(\theta,\eta) &=&se_r(k^s_n,\theta) Se_r(k^s_n,\eta), 
\end{eqnarray}
where  $ce$, $se$, $Se$ and $Ce$ are the Mathieu functions \cite{abra} . 
Of course, we have
$Se_r(k^s_n,\eta_0)=0$ and $Ce_r(k^c_n,\eta_0)=0$.  These equations
permit to find the spectrum. In figure 1 we plot a part of the spectrum
of an ellipse with $e=0.5$ and $a=71.3 \AA$.

In figure 2 we plot the LDOS at the focus in the absence of a magnetic impurity.
It is clear that only a few states of figure 1 contribute significantly
to the LDOS at the focus. The energy separation between these levels
is  much larger
than $\delta=20 meV$. There is one of these quasi-bound wavefunctions 
that has an energy of 447.5 meV,
very near $\epsilon_F$ (which, for the Cu(111) surface band is 450 meV). We can thus explain the experimental observation of a quantum
mirage in this quantum corral \cite{mirage} using the results of section III.
\begin{figure}
\centerline{
\psfig{figure=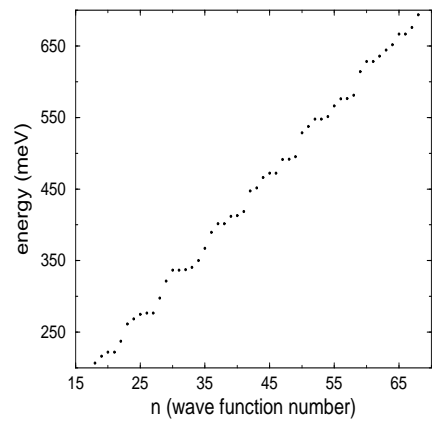,height=2.5in,width=2.5in}}
\caption{Energy spectrum of an elliptical quantum corral with $e=0.5$
and semimajor axis, $a=71.3 \AA$, on a Cu(111) surface.}
\end{figure} 
\begin{figure}
\centerline{
\psfig{figure=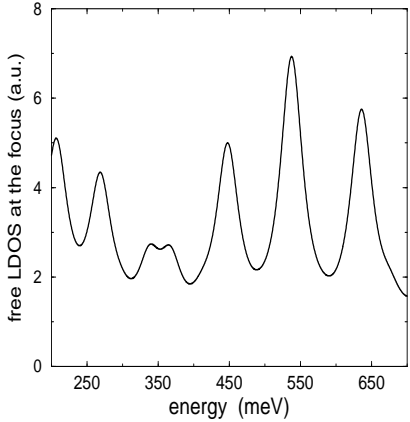,height=2.5in,width=2.5in}}
\caption{LDOS at a focus of the elliptical quantum corral with $e=0.5$,
$a=71.3 \AA$ when no magnetic impurity is present.}
\end{figure} 
In the left panel of  figure 3  we show a contour plot of the wavefunction
 at the Fermi level
for this ellipse. It must be stressed that the Fermi wavefunction  maxima are located
 at a distance of 3.28 $\AA$ of the closest focus. The lattice constant of Cu(111) is 2.55 $\AA$. Hence,
 experimentally it is very difficult to distinguish between the foci and
 the maxima. 

\begin{figure}

\centerline{
\psfig{figure=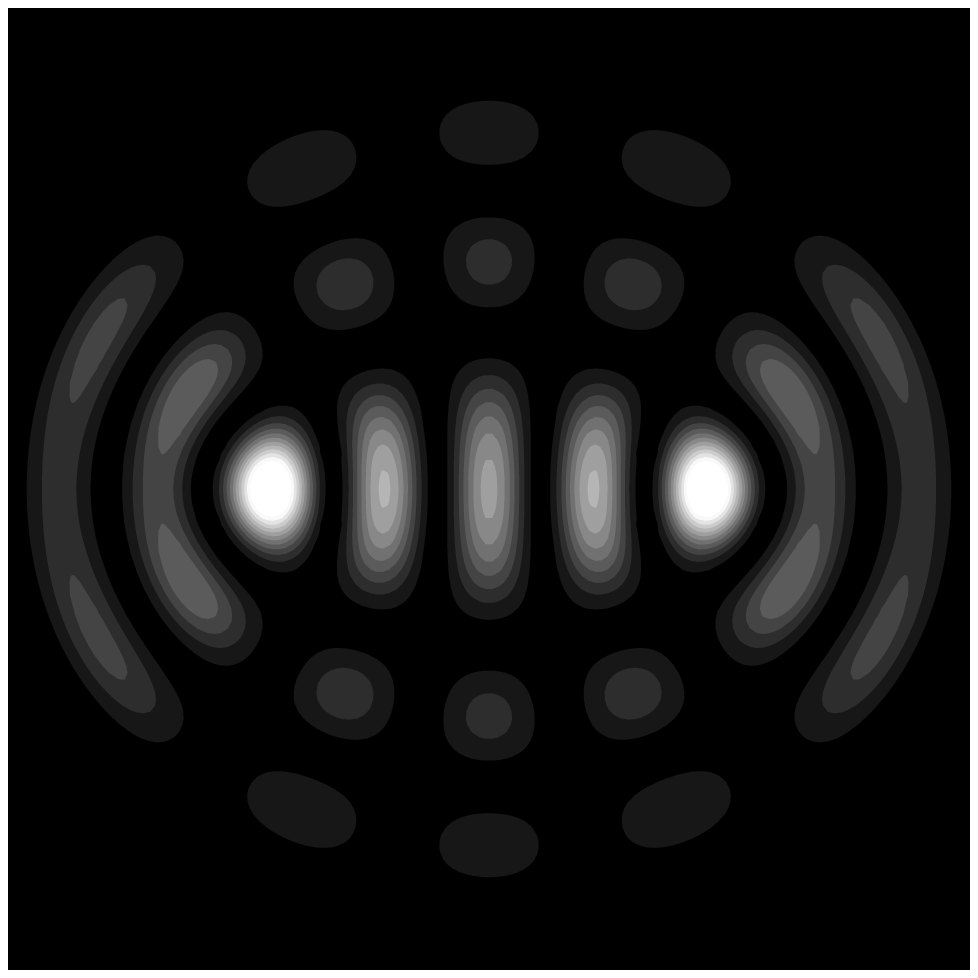,height=1.7in,width=1.7in}
\psfig{figure=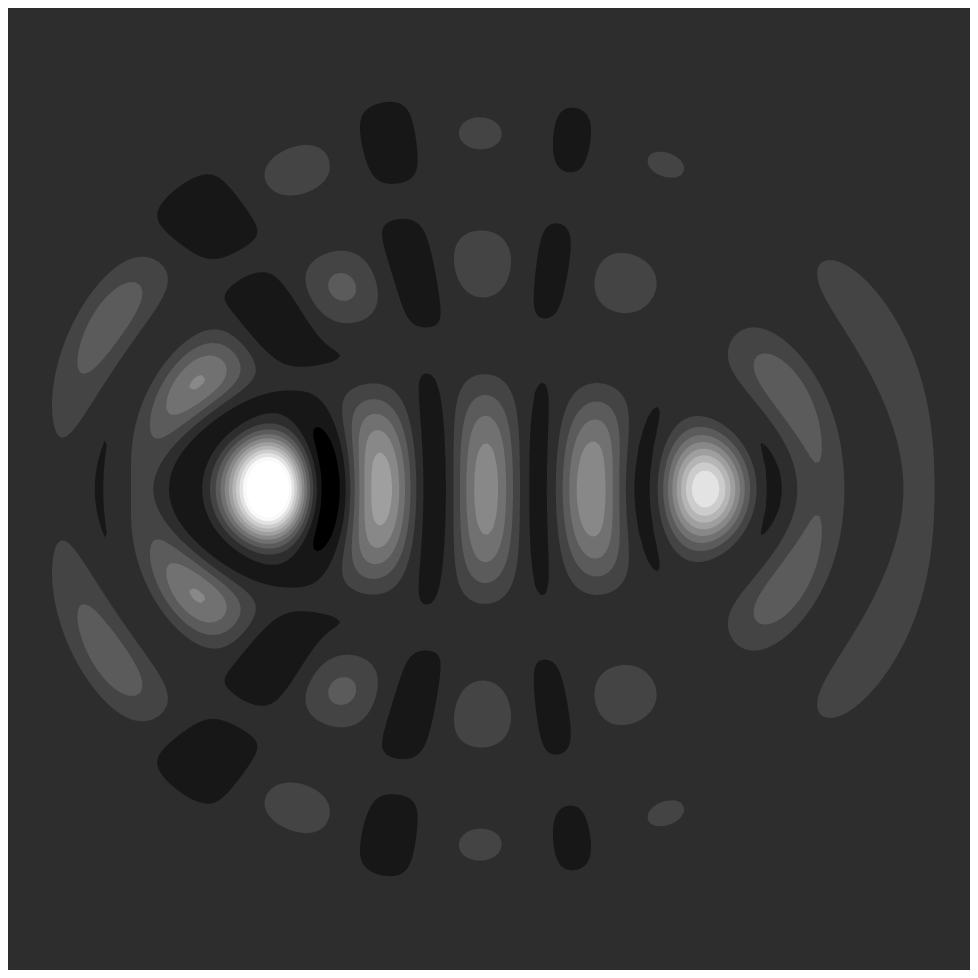,height=1.7in,width=1.7in}
}
\vspace{.1in}
\caption{Left panel: magnitude of the wavefunction at the Fermi level
for the ellipse with $a=71.3 \AA$, $e = 0.5$. 
Right panel: Change in the differential conductance due to the impurity at
the left focus, - $\delta G(V,\vec{R})$, normalized to the value in the
maximum .
Scale code: 0.75 -
1 = white, 0 = black, 0-0.75 : grey scale).}
\end{figure}

The knowledge of the corral spectrum and wavefunctions permits to
calculate $G(V,\vec{R})$ for the elliptical corral via equations
(\ref{G}), (\ref{DOS}), (\ref{GC}), (\ref{molZk}). In figure 3 (right panel),
we plot the difference between the $G(V,\vec{R})$ map with and without the 
impurity, for $eV=0$ \cite{note}.
In our calculations we take the value $k_B T_K=4.6 meV$
($T_K = 50 K$), as observed in \cite{mirage}.
In this experiment $T=4 K$, so that condition $T<<T_K$ is fulfilled.
The change in the differential conductance occurs not only at
the focus where the impurity is located but also at the empty focus, located
71 $\AA$ away from the impurity. The fingerprint of the Kondo effect is thus
dominantly located  around the impurity {\em and} around the empty focus.
 The similitude between the left and the right panel in figure 3
  supports our claim that the wavefunction of the corral at the Fermi level
   projects  the Kondo dip from the impurity to the other focus. 
   Our figure 3 should be compared with
figures 3-c and 3-d of reference \cite{mirage}.  In the case of surface
without corral, the Kondo signature would be localized around the impurity, 
being negligible at a distance larger than 2$k_F^{-1}$ \cite{Hewson}.

In figure 4 (left panel) we plot $G(V)$ when the
tip is on top of the focus where the impurity is located
(compare to figure 4(a) of reference \cite{mirage}). For this
calculation we use 
$q \approx 0.2$, the value used to fit $G(V)$ without corral 
\cite{mirage,private}, and consider a nonzero value for $t_d$, following the method
outlined in section III. In the
right panel of our figure 4 we plot $G(V)$ measured on
top of the {\em empty} focus. Hence, our theory is in agreement with
the main experimental result: the existence of a  Kondo resonance at the
empty focus, more than 80 $\AA$ away from the magnetic impurity.  It
must be stressed, however,  that in our model the dip observed on top of the
magnetic impurity and the one observed on top of the empty focus have different
lineshapes, and there exists a factor of 2 between their intensities.
In the experiment the attenuation factor is approximately $8$ and
both the original dip and the ghost are more symmetric. In order to remove
this discrepancy a less phenomenological theory for  inelastic
processes, like scattering from surface states to bulk
states, would be necessary. We predict also that
combinations of surface and adatoms for which inelastic scattering is
smaller than for Co and Cu would increase the size of the mirage.
In figure 5 we also show $G(V)$ when the tip is not at 
a maximum of the Fermi corral eigenstate. In those situations the mirage is 
not present, in agreement with the experiments \cite{mirage}.
\begin{figure}
\psfig{figure=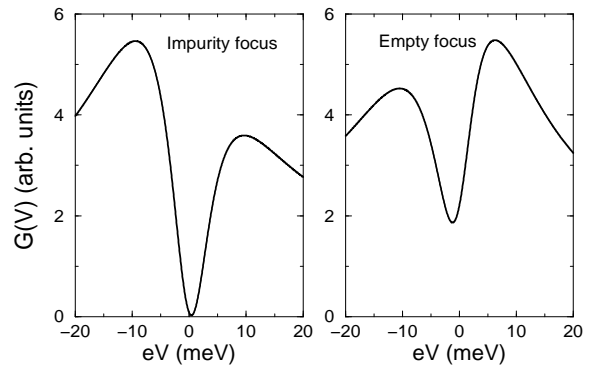,height=2.1in,width=3.5in}
\caption{Calculated dips at the focus with a magnetic impurity and at the empty
focus of the elliptical corral of figure 3.}
\end{figure} 
\begin{figure}
\psfig{figure=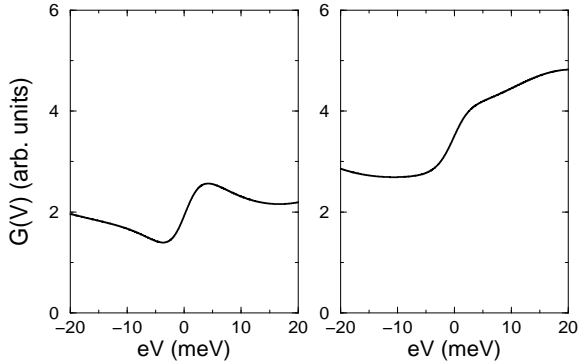,height=2.1in,width=3.5in}
\caption{$G(V)$ at the points
$\vec{R}=(25,0) \AA$ (left panel) and $\vec{R}=(35.7,25)$ \AA (right panel)
of the elliptical quantum corral of figure 3, when a magnetic impurity is placed at
the left focus. We take the center of the ellipse as the origin of coordinates.}
\end{figure}
In figure 6 we plot the intensity of the mirage (the dip amplitude)
as a function of $a$,
keeping $e=0.5$. In reference \cite{mirage}
an oscillatory dependence of the mirage effect as a  function of
$a$ (for fixed $e$) was mentioned, the oscillation 
period being $\lambda_F / 4$.
Our calculation is consistent with that claim. However, we obtain
a curve with more structure. The Fourier
transform of the intensity of the mirage shows several peaks, the largest
of which is located at $\lambda_F /4$, in agreement with the experiment.
In figure 6 we also plot the number of occupied states inside
the corral, as a function of $a$, and keeping $\epsilon_F$ constant
at 450 meV. We see that most of the changes in the occupation number
do not lead to large changes in the mirage strength. 
The mirage is only enhanced when a particular kind
of states, whose wavefunction is heavily peaked very close to the foci,
is occupied. This rule was also observed in the experiments \cite{mirage}. 
   
For the ellipse with $e=0.786$ in \cite{mirage},
we have been able to reproduce
all the results obtained for the ellipse with $e=0.5$, assuming that the
Fermi Level is somewhat below 450 meV. This indicates that the position
of the resonances given by the hard wall ellipse might not coincide with
the experimental results.

Since the maxima of the Fermi wavefunction
are not exactly located at the foci, it is our contention that the important
issue is to place the impurity at the maximum of the Fermi wavefunction. 
Therefore,
geometrical or semiclassical interpretations of the mirage might not be adequate
to address this phenomenon. To check this, we have studied a square corral,
   obtaining the mirage effect.  Elliptical corrals 
   are very convenient because 
some states with high quantum numbers (such as  the state at the Fermi level
for the $e=0.5$ ellipse with the adequate $a$) have two main maxima located close to the
   foci of the ellipse. In contrast, all the maxima in a  square corral have
   the same height so that the projection effect is less pronounced than
   in the ellipse. 
\begin{figure}
\psfig{figure=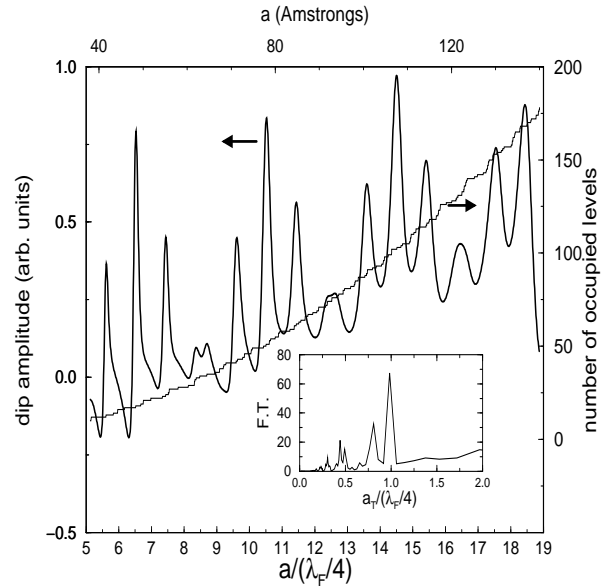,height=3.in,width=3.in}
\caption{Thick line: Dip amplitude as a function of $a$, $dip(a)$, 
for an elliptical corral
with $e=0.5$. We display $a$ in $\AA$ (upper axis) and in $\lambda_F /4$
units (lower axis). Thin line: number of occupied quasi-bound wavefunctions
inside the ellipse. In the inset we display the Fourier Transform 
$f(a_T)\propto \int_{0} ^{\infty} dip(a) \; exp(-i \ 2 \pi \ a/a_T) \;da$ of
this curve as a function of the period. We see that the curve peaks at
$a_T/\frac{\lambda_F}{4} =1$. }
\end{figure}

\section{Discussion and conclusions}

We now comment on some of the limitations of our theory. The first has to do
with the approximation of the eigenstates inside the quantum corral
as quasi-bound states broadened in energy.
For a quantitative description of the corral energy spectrum a
more detailed calculation is needed, taking into account the role of the corral 
atoms as tunneling centers to the bulk states \cite{Heller}.
The second is the use of the Friedel sum rule in a resonant level model.
A more realistic calculation of the
impurity Green's function would imply to take into account the real wavefunctions
inside the corral and the possibility of tunneling from the magnetic impurity to
the bulk states. The quantitative discrepancy with the experiments, in what
concerns the attenuation of the mirage, should be solved including these
effects. A more complete theory for the STM through magnetic impurities in metallic
surfaces without quantum corrals has been developed in
\cite{Schiller,Ujs,Ujsreview}.

The emphasis  of this paper is placed on the qualitative understanding
of the mirage rather than on a detailed description
of the experiments. Our main results are the following:
 {\em  i)} The LDOS evaluated at an arbitrary surface point,
 $\vec{R}$, in the Anderson model, 
contains information about the  LDOS at the impurity site, $\vec{R}_I$.
 A mirage will appear in a remote point, $\vec{R}$,  whenever  there
 is a single quantum state at the Fermi level whose amplitude
  $\psi_{\epsilon_F}$  peaks  both at the impurity ($\vec{R}_I$) 
   and at $\vec{R}$. In order to avoid destructive interference between 
   different
   states it is necessary that the energy spacing between states
   with a non-negligible amplitude at the impurity site is bigger than the
   energy broadening, $\delta$.  
{\em ii)}  The mirage can be obtained in 
corrals with shapes other than elliptical. However, the elliptical shape
is quite convenient because some of the corral eigenstates   peak strongly
at two points very close to  the foci.
{\em iii)} Our theory predicts that the intensity of the 
mirage in an elliptical corral oscillates,
 as a function of the semimajor axis length,
keeping fixed
the eccentricity, with a dominant period of $\lambda_F/4$, in agreement
with reference \cite{mirage}.

We want to acknowledge C. Piermarocchi and H. Manoharan for fruitful discussions.
JFR and DP acknowledge  Spanish Ministry of Education (MEC)  for postdoctoral
research fellowship and grant FPU/AP98. Work supported in part by MEC under contract
PB96-0085.

 During the completion of this manuscript we became aware of a theoretical
 work addressing the problem of the mirage in an elliptical quantum corral 
 \cite{Schiller2}. In that work the states of the ellipse are described by
 a more detailed method, assuming that the wall atoms are magnetic, and the
 issue of the existence of the mirage in different geometries is not addressed.
 Our theory
 can be applied to the general case of a quantum corral formed by non-magnetic
 scatterers (in which quantum mirages have also been observed \cite{mirage}).

\widetext


\begin{references}
\bibitem{note} We have also repeated our calculations considering tunneling from the
STM tip to a finite surface in the corral (instead of tunneling to a point).
We have
obtained very similar results for both cases.
\end{references}
\end{document}